\documentclass[%
 aip,
 amsmath,amssymb,
 reprint,%
]{revtex4-1}

\usepackage{graphicx}
\usepackage{dcolumn}
\usepackage{bm}

\usepackage{graphicx}
\usepackage{caption}
\usepackage{subcaption}

\usepackage[utf8]{inputenc}
\usepackage[T1]{fontenc}
\usepackage{mathptmx}
\usepackage{etoolbox}
\usepackage{soul}

\makeatletter
\def\@email#1#2{%
 \endgroup
 \patchcmd{\titleblock@produce}
  {\frontmatter@RRAPformat}
  {\frontmatter@RRAPformat{\produce@RRAP{*#1\href{mailto:#2}{#2}}}\frontmatter@RRAPformat}
  {}{}
}%
\makeatother

\usepackage{xcolor}
\usepackage{dsfont}

\usepackage{graphicx}
\usepackage{hyperref}
\newcommand{\ben}{\begin{equation}}
\newcommand{\een}{\end{equation}}
\newcommand{\bea}{\begin{eqnarray}}
\newcommand{\eea}{\end{eqnarray}}
\def\bra#1{\langle#1\vert}
\def\ket#1{\vert#1\rangle}

\def\sss{\scriptscriptstyle\rm}

\def\1s{_{1,\sss S}}
\def\2s{_{2,\sss S}}

\def\s{_{\sss S}}
\def\xc{_{\sss XC}}

\def\Hxc{_{\sss HXC}}

\def\ext{_{\rm ext}}

\def\Hxcqq{_{{\sss HXC}\,qq}}
\def\Hxcqqp{_{{\sss HXC}\,qq'}}

\def\br{{\bf r}}

\begin{document}

\preprint{AIP/123-QED}

\title{Oscillator strengths and excited-state couplings for double excitations in time-dependent density functional theory}
\author{Davood B. Dar}
\author{Neepa T. Maitra}%
 \email{neepa.maitra@rutgers.edu}
\affiliation{ Department of Physics, Rutgers University, Newark 07102, New Jersey USA
}%

\date{\today}

\begin{abstract}
Although useful to extract excitation energies of states of double-excitation character in time-dependent density functional theory that are missing in the adiabatic approximation,  the frequency-dependent kernel derived earlier [J. Chem. Phys. {\bf 120}, 5932 (2004)] was not designed to yield oscillator strengths. These are required to fully determine linear absorption spectra and they also impact excited-to-excited-state couplings that appear in dynamics simulations and other quadratic response properties. Here we derive a modified non-adiabatic kernel that yields both accurate excitation energies and oscillator strengths for these states. We demonstrate its performance on a model two-electron system,  the Be atom, and on excited-state transition dipoles in the  LiH molecule at stretched bond-lengths, in all cases producing significant improvements over the traditional approximations.
\end{abstract}

\maketitle

The challenge of describing excited states of double-excitation character has grown from being a relatively minor nuisance in certain regions of linear response spectra to being a real hindrance in our ability to simulate photo-excited dynamics in molecules. 
These states are loosely defined in the sense of having a significant proportion of doubly-excited determinants with respect to a reference ground-state Slater determinant, where a doubly-excited determinant promotes
two occupied orbitals of the reference determinant to two virtual ones~\cite{M22}. 

In ground-state spectra these states have typically smaller transition strengths (``darker") than predominantly singly-excited states but they can siphon off oscillator strength of neighbouring single-excitations whose absorption peaks will be overestimated in simulations that do not account for double-excitations. Moreover, these states are readily accessible when a molecule is excited, offering photochemical/physical pathways that play a role in a range of applications, from photocatalytic design, photo-induced charge transport, to laser-driven or polariton-enabled control. 
While the true correlated excited-state wavefunctions contain contributions from double excitations, these are either absent in approximate electronic structure methods or poorly approximated, resulting in states  missing from the spectrum, or with energies and transition strengths with larger errors than for states with predominantly single-excitation character~\cite{LBSCJ19}, wreaking havoc on predictions of excited-state dynamics.

The problem is particularly severe for single-reference methods in their usual modus operandi, such as time-dependent density functional theory (TDDFT) within the adiabatic approximation~\cite{JCS96,TH00,MZCB04,M22}, equation-of-motion coupled-cluster (EOM-CC) with singles and doubles only~\cite{SNB00,LBSCJ19}, or the Bethe-Salpeter equation with the GW approximation (GW-BSE) with static screening~\cite{RSBS09,SROM11,BB22}. While double-excitation character can be adequately included in EOM-CC if triple excitations are included in the cluster expansion, and in GW-BSE with full frequency-dependence or equivalently in an expanded excitation space, these procedures increase the computational cost of these methods, both of which are already more expensive than TDDFT.  Multiconfigurational methods may be better platforms for these states, but aside from their higher computational cost, other issues enter such as gauge-sensitivity, and the sizes of an adequate active space which can change during a dynamical process.  For determination of oscillator strengths for these states (or any excited state) the difficulty increases significantly since increased accuracy 
requires cranking up the number of electronic configurations more so than for corresponding improvement in energies, which results in a large increase in the computational cost\cite{DQCJKSL23,SBMLJ21}.

Fixing adiabatic TDDFT thus becomes an attractive option. Indeed, double-excitations provide a prime example where TDDFT in theory and in practise diverge: in theory, their excitation energies and oscillator strengths are captured exactly, but the approximate functionals used in practise cannot capture them. It is now well-known that the culprit is the adiabatic nature of the approximation~\cite{JCS96,TH00, MZCB04,M22}: the exact exchange-correlation (xc) kernel has a strong frequency-dependence (a pole) that effectively folds a double-excitation of the non-interacting Kohn-Sham (KS) system into the TDDFT response equations, but adiabatic approximations use a frequency-independent kernel. The frequency-dependence implicit in orbital-dependence of a hybrid functional does not have the correct structure. Using first-principles arguments, Ref.~\cite{MZCB04} proposed a frequency-dependent kernel, sometimes called ``dressed TDDFT", that has been shown to successfully capture the energies of double-excitations in a range of molecules~\cite{CZMB04,EGCM11,HIRC11,MW09,MMWA11}. However,  it is unable to yield information on the oscillator strengths because it is designed to operate within the framework of a Tamm-Dancoff-like approximation which does not preserve oscillator strengths. 

In this Communication, we derive a kernel designed to work within full TDDFT linear response, which yields both excitation energies and oscillator strengths of these states, preserving the oscillator strength sum-rule.
We demonstrate its accuracy on a two-electron model system, the lowest $^1D$ excitations in the Be atom, and the LiH molecule over a range of interatomic separations. The kernel provides an approximation of how the transition density of a KS single excitation gets distributed into  mixed single- and double-excitations of the true system. This enters into a formula to compute excited-to-excited state transition densities that was recently derived~\cite{DRM23} to tame unphysical divergences in the adiabatic quadratic response function giving an improvement in the transition dipole between two states in the LiH molecule as it begins to dissociate, with accuracy similar to that of the more {\it ad hoc} pseudowavefunction approximation~\cite{LL14,ZH15,OAS15,OBFS15,PRF16}.

In TDDFT, excitation spectra are extracted from the linear response of the density,  usually cast in the form of a matrix equation in the space of single KS excitations, $q = i \to a$, with $i (a)$ labelling occupied (unoccupied) orbitals~\cite{C95,PGG96,GPG00,GM12chap}:
\ben
\Omega(\omega){\bf G}_I = \omega_I^2 {\bf G}_I
\label{eq:casida}
\een
where 
\ben
\Omega(\omega) = \nu^2_q \delta_{qq'} + 4 \sqrt{\nu_q\nu_q'} f\Hxcqqp(\omega)\,.
\label{eq:casida_mat}
\een
Here $f\Hxcqqp=\int d\br d\br' \Phi_q(\br) f\Hxc(\br,\br',\omega) \Phi_{q'}(\br')$ is the matrix element of the Hartree-exchange-correlation kernel,
$f\Hxc(\br,\br',\omega)=\frac{1}{\vert \br-\br' \vert}+f\xc(\br ,\br' ,\omega)$ with $\Phi_q =\phi_i^* \phi_a$, being the product of occupied and unoccupied orbitals. The KS frequencies $\nu_q = \epsilon_a - \epsilon_i$ are corrected to the true ones $\omega_I$ through solving Eq.~(\ref{eq:casida}).
While the square-root of the (pseudo-)eigenvalues, $\omega_I$, give the excitation frequencies of the physical system, the oscillator strength of the transition can be obtained from ${\bf G}_I$ when normalized according to~\cite{C95}
\ben
{\bf G}_I^\dagger\left(\mathds{1} - \left.\frac{d\Omega}{d\omega^2}\right\vert_{\omega  =\omega_I}\right){\bf G}_I = 1\,.
\label{eq:casida_os}
\een
Oscillator strengths $f_I$ (or any one-body transition properties) can be obtained from these eigenvectors, through
\ben
f_I \equiv \frac{2}{3}\omega_I \vert\bra{\Psi_0}\hat{\br}\ket{\Psi_I}\vert^2
=\frac{2}{3} \left\vert \sum_{q,q'} \br^{\dagger}_q S_{q,q'}^{-1/2}{G}_{I,q'}\right\vert^2
\label{eq:osc_str}
\een
where on the right $\br = (x, y, z)$   with $x$  being a vector in single-excitation space with matrix elements $x_q = \bra{\phi_i}x\ket{\phi_a}$ and $S$ is a diagonal matrix $S_{qq} = \frac{1}{\nu_q} = \frac{1}{\epsilon_a -\epsilon_i}$. 
The oscillator strength sum-rule, $\sum_I f_I = N$ is satisfied by both the true and KS systems and Eq.~(\ref{eq:osc_str}) gives the contributions of the KS oscillator strengths to a given true $f_I$ through the normalized eigenvector ${\bf G}_I$.
In the adiabatic approximation $\Omega(\omega) = \Omega$ has no frequency-dependence so the (pseudo-)eigenvectors are just 
normalized to $1$, but a frequency-dependent kernel causes a redistribution of oscillator strengths through Eq.~(\ref{eq:casida_os}).

A key point is that, whether with a frequency-dependent kernel or not, accurate oscillator strengths fulfilling the oscillator strength sum-rule  can only be obtained when the full TDDFT matrix Eq.~(\ref{eq:casida_mat}) is used, i.e. including both backward and forward transitions~\cite{GM12chap,HH99}. Backward transitions are neglected in the Tamm-Dancoff approximation where eigenvalues of the corresponding matrix are directly the frequencies $\omega_I$
(rather than $\omega_I^2$).
Although this approximation has been argued to sometimes give more accurate excitation energies with some approximate functionals~\cite{LFHH22,CGGG00,HH99}, the oscillator strengths are generally worse, and the sum-rule violated.

Eq.~(\ref{eq:casida_os}), derived in Ref.~\cite{C95}, does not appear to be well known and, to our knowledge, used only once before~\cite{CFMB18} in a detailed study of the linear response of the Hubbard dimer. In general, there has been much more attention paid to excitation energies than to oscillator strengths and the latter have only been studied with the adiabatic approximation~\cite{SBMLJ21,CBLJ21,R18b,CTFW11,LJ13} with eigenvectors normalized to 1. Most interesting for present purposes, Eq.~(\ref{eq:casida_os}) tells us how the oscillator strength of a KS single excitation that lies near a KS double excitation shatters into oscillator strengths of the mixed single and double excitations, when an appropriate frequency-dependent kernel is used. 

 Consider the limit that a KS excitation is isolated enough in energy from neighbouring excitations, such that the diagonal corrections in the TDDFT matrix dominates, and the full TDDFT matrix Eq.~(\ref{eq:casida_mat}) locally reduces to the $1\times 1$ ``small-matrix approximation (SMA)", 
\ben
\Omega(\omega) = \nu_q^2 + 4\nu_q f\Hxcqq(\omega)
\label{eq:sma}
\een
Setting this equal to $\omega^2$, we observe that frequency-dependence of the kernel $f\xc(\br, \br',\omega)$ creates more than one TDDFT excitation from this single KS excitation. The eigenvectors $G_I$ reduce to one number for each eigenvalue, and indicate the fraction of the KS oscillator strength that goes into the corresponding true transition, with the second equality in Eq.~(\ref{eq:osc_str}) reducing to
\ben
|G_{I}|^2 = \frac{\omega_I \vert \langle \Psi_0 \vert \br \vert \Psi_I\rangle\vert^2}{\nu_q \vert \langle \Phi_0 \vert \br \vert \Phi_q\rangle\vert^2}\,.
\label{eq:G-os}
\een
In this case, the oscillator strength associated  with this isolated KS transition will be preserved over the true excitations in this subspace provided that the approximation for $f\Hxc(\omega)$ that will go into $\Omega(\omega)$
gives eigenvectors with the property $\sum_I G_I^2 = 1$: 
\ben
\sum_I G_I^2 = 1\;\; \Rightarrow \;\;\nu_q \vert \bra{\Phi_0} x\ket{\Phi_q}\vert^2 = \sum_{I}\omega_{I} \vert \bra{\Psi_0} x\ket{\Psi_{I}}\vert^2
\label{eq:trunc_sumrule}
\een
where $I$ labels the excitations generated from the single KS excitation by the frequency-dependent kernel. This is an important condition for the approximation for $f\Hxc(\omega)$. We note that it is satisfied for any $\Omega(\omega)$ of the form 
\ben
\Omega(\omega) = A + B/(\omega^2 -D)
\label{eq:goodform}
\een
where $A, B, D$ are constants.

We further note that since the matrix elements in Eq.~\ref{eq:osc_str} (Eq.~\ref{eq:G-os}) involve a one-body operator, then in the limit that the true ground-state is dominated by a single Slater determinant (weak interaction limit),  only the underlying single-excitation components of $\Psi_I$ contribute to $\vert G_I\vert^2$. Since $\sum_I G_I^2 = 1$, this means that $1 - \vert G_I\vert^2$ contains the double-excitation (and any higher-excitation) contribution, in this limit.

While the ``dressed TDDFT" kernel derived in Ref.~\cite{MZCB04} has the correct form of frequency-dependence to yield energies of double-excitations, it is based on the Tamm-Dancoff approximation, which reduces to the single-pole approximation (SPA) in the isolated KS excitation case, and cannot be used in a consistent way to obtain oscillator strengths. Instead, here we follow a similar approach to Ref.~\cite{MZCB04} but within the SMA. We begin by writing the many-body time-independent Schr\"odinger equation such that square of the frequency is the eigenvalue: $(H - E_0)^2\Psi = \omega^2 \Psi$, evaluating this eigenvalue problem in the KS basis, and working in the limit where a KS single-excitation $q$ and double-excitation $d$ are well-separated in energy from all the other excitations of the system. We define $\nu_d = \nu_{s1} + \nu_{s2}$ as the sum of the KS single-excitations $\nu_{s1}$ and $\nu_{s2}$ such that $\nu_d$ is close to $\nu_q$. Diagonalization then yields
\ben
\omega^2 = (H_{qq} - E_0)^2 + \vert H_{qd}\vert^2 \left(1 + \frac{(H_{qq}+H_{dd}-2E_{0})^{2}}{\left[\omega^{2}-\left((H_{dd}-E_{0})^{2}+H_{qd}^{2}\right)\right]}\right) 
\label{eq:raw}
\een
where $H_{ij}$ are the matrix elements of the interacting Hamiltonian in the basis $\{q,d \}$ which spans the truncated Hilbert space containing the single and the double excitations, and $E_0$ is the ground state energy.
However Eq.~(\ref{eq:raw}) is nothing but exact diagonalization in the KS basis; to take advantage of the correlation contained in ground-state TDDFT approximations and to extract a frequency-dependent kernel for an improved TDDFT approximation, we replace $H_{qq} - E_0$ with its adiabatic SMA value, and identify the remainder as defining a frequency-dependent kernel:
\bea
\omega^2 &=& \nu_q^2 + 4 \nu_q f\Hxcqq^{\rm DSMA_0}(\omega)\,,\; {\rm where}\\
f\Hxcqq^{\rm DSMA_0} (\omega) &=& f\Hxcqq^{\rm A} + \frac{\vert H_{qd}\vert^2}{4\nu_q} \left(1 + \frac{(H_{qq}+H_{dd}-2H_{00})^{2}}{\left[\omega^{2}-\left((H_{dd}-H_{00})^{2}+H_{qd}^{2}\right)\right]}\right)
\label{eq:DSMA1.1}
\eea
In Eq.~(\ref{eq:DSMA1.1}) we also replaced $E_0$ in Eq.~(\ref{eq:raw}) by $H_{00}$ to better balance errors in the truncated matrix elements (as was done for the dressed SPA of Ref.~\cite{MZCB04}). 

Eq.~(\ref{eq:DSMA1.1}) is our central equation, and it becomes exact in the limit that the coupling between the single and double excitation induced by the electron-interaction is much stronger than the coupling with other single excitations. While the adiabatic approximation merely shifts the KS frequency, Eq.~(\ref{eq:DSMA1.1}) creates a new one, yielding two positive frequencies $\omega_1, \omega_2$ which approximate the mixed single- and double- excitations of the true system, similar to the dressed SPA kernel of Ref.~\cite{MZCB04}. However, unlike that kernel, Eq.(\ref{eq:DSMA1.1}) preserves the oscillator strength: 

It has the form of Eq.~\ref{eq:goodform} and using Eq.~(\ref{eq:sma}) with Eq.~(\ref{eq:DSMA1.1}),   Eq.~(\ref{eq:casida_os}) explicitly finds that the eigenvectors $G_{1(2)}$ (one number  for each eigenvalue in this $1\times 1$ case) are such that $G_1^2 + G_2^2 = 1$, thus satisfying the sum-rule (Eq.~\ref{eq:trunc_sumrule}), 
\ben
\nu_q \vert \langle \Phi_0 \vert x \vert \Phi_q\rangle\vert^2 = \omega_1 \vert \langle \Psi_0 \vert x \vert \Psi_1\rangle\vert^2 + \omega_2 \vert \langle \Psi_0 \vert x \vert \Psi_2\rangle\vert^2 
\een

The resulting $G_{1,2}$ from  Eq.~(\ref{eq:casida_os}) gives the fraction of the KS oscillator strength that goes into the respective transition $I = 1, 2$ (see Eq.~(\ref{eq:G-os})), with $1 - G_I^2$ giving the double (and higher) excitation contribution to that state in the limit that the ground-state is dominated by a single determinant (see earlier).

In the limit that the coupling between the single and double excitation $H_{qd}$ goes to zero, Eq.~\ref{eq:DSMA1.1} correctly reduces to the adiabatic approximation chosen for $f\Hxc^{\rm A}$. When $H_{qd}$ is much smaller than all the other terms, then one frequency is a slightly corrected adiabatic value, while the other reduces to $\sqrt{(H_{dd} - H_{00})^2+ H_{qd}^2}$. This motivates a variation of Eq.~(\ref{eq:DSMA1.1}) where we replace the diagonal matrix element differences  with KS values:
\ben
f\Hxcqq^{\rm DSMAs} (\omega) = f\Hxcqq^{\rm A} + \frac{\vert H_{qd}\vert^2}{4\nu_q} \left(1 + \frac{(\nu_{q}+\nu_{d})^{2}}{\left[\omega^{2}-\left(\nu_{d}^{2}+H_{qd}^{2}\right)\right]}\right)\,.
\label{eq:DSMAs}
\een
Another possibility is to replace them with their adiabatic TDDFT counterparts:
\ben
f\Hxcqq^{\rm DSMAa} (\omega) = f\Hxcqq^{\rm A} + \frac{\vert H_{qd}\vert^2}{4\nu_q} \left(1 + \frac{(\Omega_{q}^A+\Omega_{s1}^A+\Omega_{s2}^A)^{2}}{\left[\omega^{2}-\left(((\Omega_{s1}^A+\Omega_{s2}^A)^{2}+H_{qd}^{2}\right)\right]}\right)
\label{eq:DSMAa}
\een
where $\Omega_{q,s1,s2}^A$ are the adiabatic TDDFT frequencies that correct the bare KS frequencies $\nu_{q,s1,s2}$.
Eq.~(\ref{eq:DSMAs}) and~(\ref{eq:DSMAa}) have a numerical advantage in requiring less two-electron integrals than in Eq.~(\ref{eq:DSMA1.1}) but all three flavours of DSMA capture excitation energies and oscillator strengths of double excitations as we will now demonstrate on explicit examples. 

\begin{figure}[h]
\includegraphics[width=0.5\textwidth]{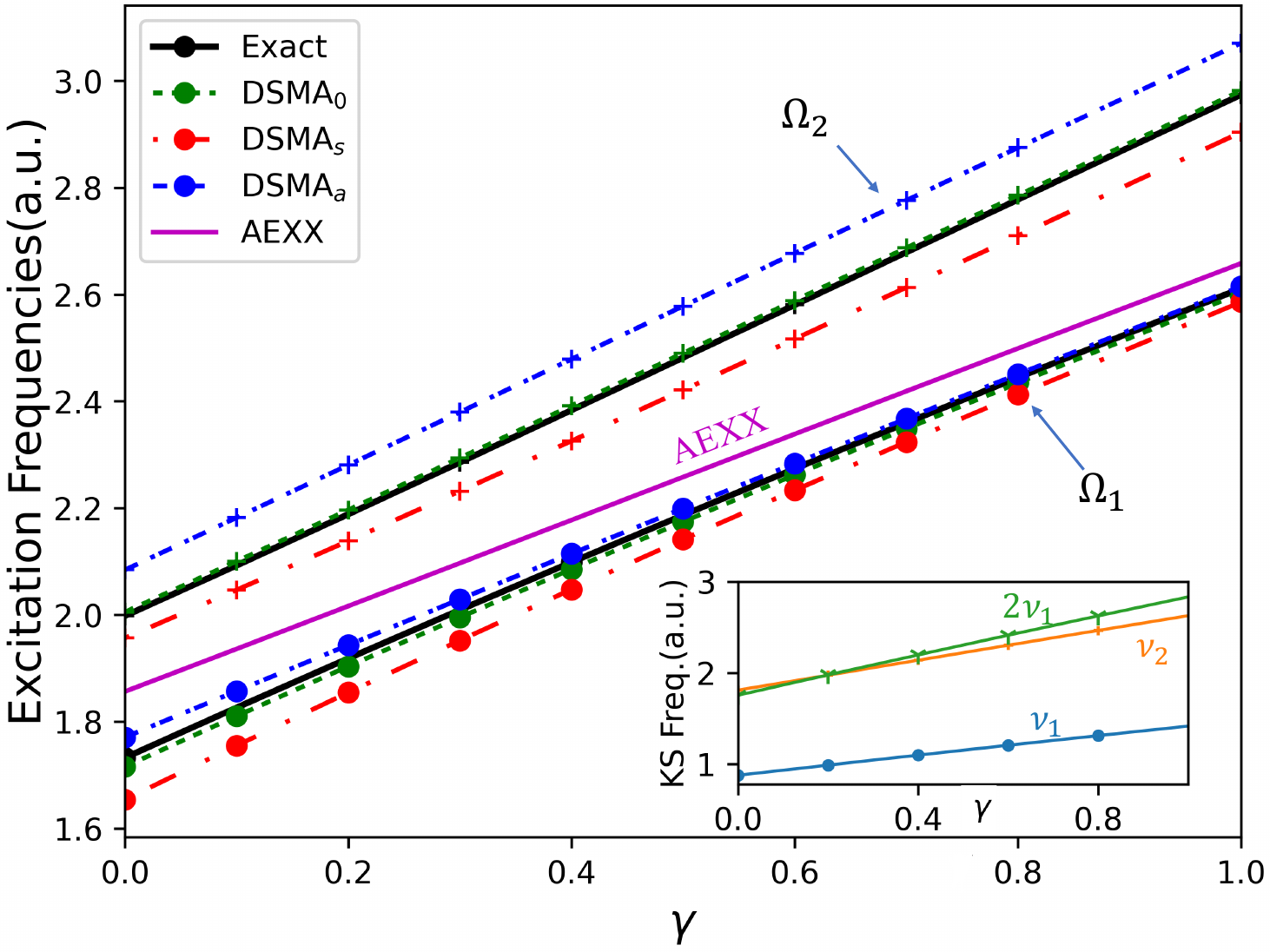} 
\includegraphics[width=0.5\textwidth]{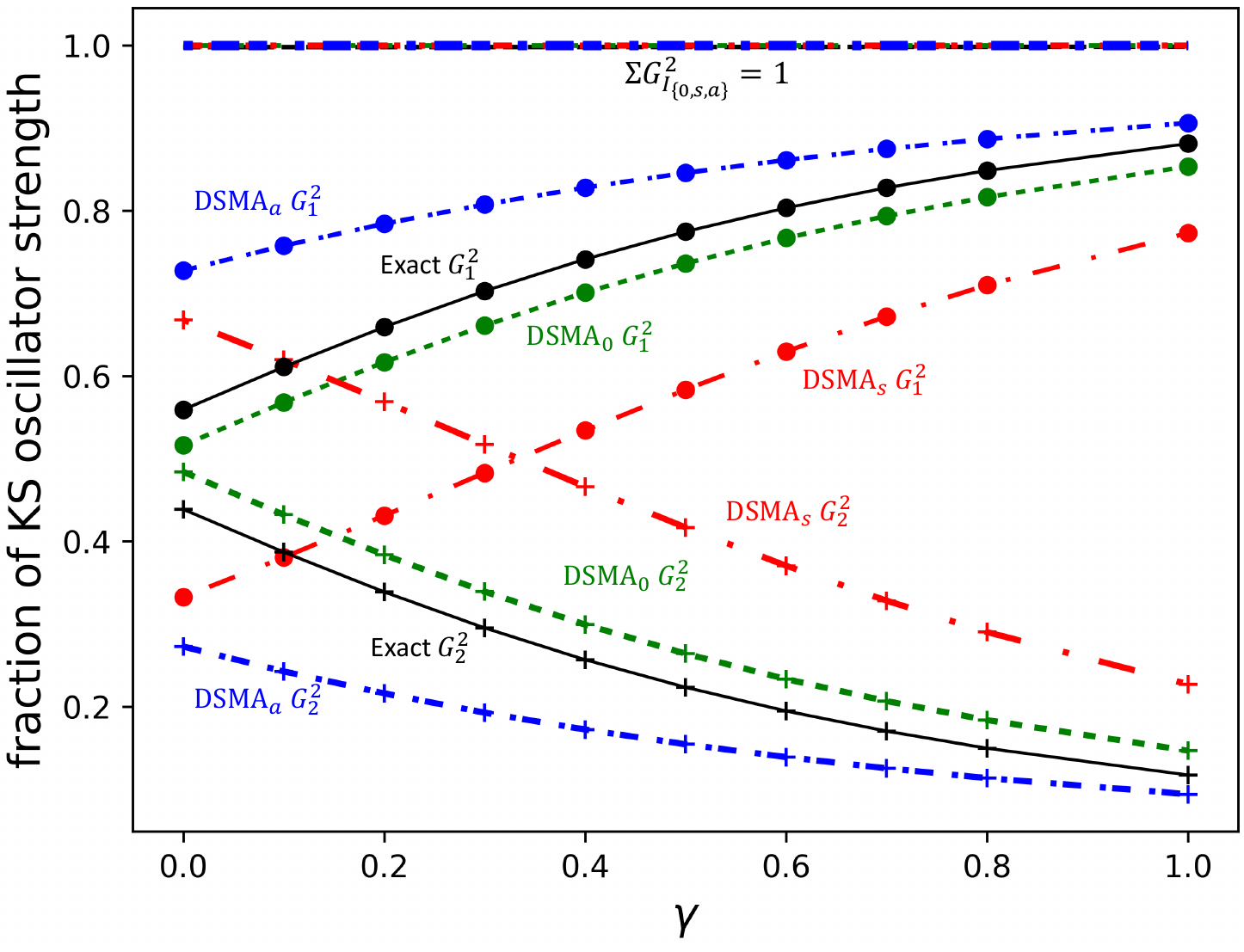}
\caption{Two electron model: a)  Exact, AEXX, and the three flavors of DSMA frequencies in the second multiplet, as a function of $\gamma$. The inset shows the lowest KS excitation frequencies.
b) The fraction of KS oscillator strengths, $G_I^2$ in the exact and the three DSMA calculations. Also shown is the sum-rule $G_1^2 + G_2^2 = 1$. }
\label{fig:model}
\end{figure}

Figure~\ref{fig:model} shows the performance on an exactly-solvable model system: two electrons in a one-dimensional harmonic plus linear potential, $v\ext(x) = \frac{1}{2}x^2 + \gamma \vert x\vert$ where $\gamma$ is a parameter in the range$[0,1]$; varying $\gamma$ tunes the degree of mixing of the single and double-excitation character in the interacting excited states. The electrons interact via a soft-Coulomb interaction: $\frac{1}{\sqrt{(x_1-x_2)^2+1}}$. 
We observe in Fig.~\ref{fig:model}(a) that for small enough $\gamma$ the exact KS system has a double excitation  lying very close to the second KS single excitation, with which it mixes strongly when the interaction is turned on as in the physical system, producing two states. As expected, the adiabatic approximation, here chosen to be exact exchange (AEXX),  is blind to the double excitation and has only one excitation in this region of the spectrum, while all DSMA approximations correctly yield two excitations. (Note that the exact ground-state KS potential is used to find the bare KS orbitals and energies, while AEXX is used for $f\Hxc^{\rm A}$). Out of the DSMA's outlined above, DSMA$_0$ performs the best, giving excitation frequencies very close to the exact. 
The excitation energies provided by each variant of DSMA are close to those given by the corresponding version of DSPA (see the Supplementary Material). However, the real and significant improvement of DSMA over DSPA lies in its capacity to accurately determine oscillator strengths.
Fig.~\ref{fig:model}(b) shows the fractions of oscillator strength shared by the two states. Again we see that DSMA$_0$ is closest to the exact, but all give a significant improvement over the adiabatic result of 1 and 0 for any $\gamma$ and the DSPA fractions shown in the Supplementary Material.  All three flavors of DSMA respect the oscillator strength sum rule, Eq.~(\ref{eq:G-os}) as shown in the figure, in contrast to the DSPA (see Supplementary Material for the analogous plot).

\begin{figure}[h]
\includegraphics[width=0.5\textwidth,height=0.2\textheight]{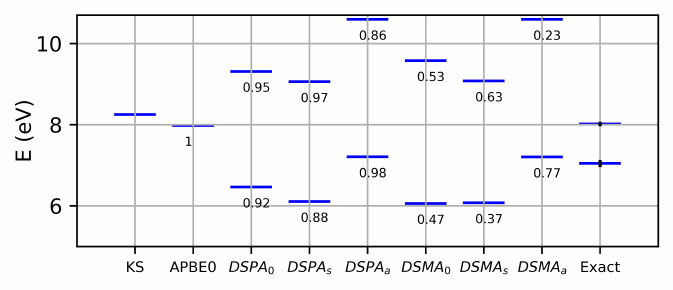} 
\caption{Be atom $1 ^1D$ excitation frequencies marked with their fraction of KS excitation oscillator strength, $G_I^2$ of Eq.~\ref{eq:G-os}}
\label{fig:Be}
\end{figure}

We next turn our attention to the Beryllium atom. Here the lowest two singlet $^1D$ states, have a mixed single and double excitation character. Ref.~\cite{LBSCJ19} reports a roughly $30\%$ single-excitation character to the predominantly doubly-excited  $ 1^1D$ state ($1 ^1S (2s^2) \rightarrow 1 ^1D(2p^2)$. As a result, adiabatic TDDFT fails to describe the states accurately. Figure~\ref{fig:Be} shows a plot of the two excitation frequencies of the $1 ^1D$ multiplet. Adiabatic TDDFT with the PBE functional gives only one state in this frequency region which is closer in energy to the upper exact state, which would take up all the oscillator strength ($G_q^2=1$) within the SMA.  The three flavors of DSMA give two states, but overestimate their energy splitting. Still, DSMA$_0$ and especially DSMA$\s$ yield oscillator strengths which  are close to that of the reference Ref.~\cite{LBSCJ19} for the lower excitation, within the assumption that Be is weakly enough correlated that 
 $G_I^2$ meaningfully approximates the single-excitation component of the state $I$. On the other hand, the three flavors of DSPA also give two excitation frequencies but yield wrong oscillator strengths and violate the oscillator sum rule.

Finally we turn to the LiH molecule at stretched bond lengths, which we will use to illustrate the use of our DSMA to calculate coupling matrix elements between excited states. Adiabatic approximations to calculate these  within quadratic response are plagued with  unphysical divergences that appear when the difference between two excitation energies $\Omega_a$, and  $\Omega_c$ coincides with another excitation frequency $\Omega_b = \Omega_c - \Omega_a$~\cite{PRF16,LL14,ZH15, OBFS15} (see Supplementary Information for a figure of this). 
In Ref.~\cite{DRM23} a frequency-dependent approximation for the second-order response kernel, $g\xc^{\rm App}$, was derived in a truncated Hilbert space containing these states, that eliminated these divergences. This kernel was used to derive the transition density $n_{ca}(\br) = \bra{\Psi_c}\hat{n}(\br)\ket{\Psi_a}$ for a case where $\Omega_a \approx \Omega_b  \approx \Omega_c/2$:

\bea
\begin{aligned}
n_{ca}^{\rm App}(\br) &= \sqrt{1 - G_{c}^2}  n_{{\sss S}1d}(\br) \\
&\quad +  \sqrt{\left(\frac{\nu_3\nu_1^3}{\Omega_c\Omega_a^3}\right)}G_c\left(1 - 2\frac{\langle f\Hxc(\Omega_a) \rangle_1 }{\Omega_a}\right)n_{{\sss S},13}(\br) \\
&\quad - \frac{n_{0a}(\br)}{\Omega_a}\int n_{0c}(r_1)g\xc^{\rm adia}(\br_1,\br_2,\br_3) n_{0a}(\br_2)n_{0a}(\br_3)d\br_1 d\br_2 d\br_3
\end{aligned}
\label{eq:approx-nac}
\eea
Here $\nu_1$ and $\nu_3$ are the frequencies of the dominant KS states contributing to the interacting levels $\Omega_a$ and $\Omega_c$ respectively, $\langle f\Hxc(\Omega_a)\rangle_1 = \int \phi_0(\br)\phi_1(\br) f\Hxc(\br,\br',\Omega_a)\phi_0(\br')\phi_1(\br') d^3\br d^3\br'$. In the case considered, state $a$ has predominantly single-excitation character, and a frequency-independent kernel suffices for this kernel. The third term involves  $g^{\rm adia}_{{\sss XC}}(\br_1,\br_2,\br_3) = \left.\frac{\delta^3 E\xc[n]}{\delta n(\br_1) \delta n(\br_2) \delta n(\br_3)}\right\vert_{n = n_0}$, an adiabatic approximation to the second-order response kernel.
The first term is the contribution from a double excitation, which should be weighted by the doubles contribution to state $c$, which, in the limit of weak enough interaction is $\sqrt{1 - \vert G_{c}\vert^2}$, however until the present work, how to calculate this was unknown. We note that Eq.~(\ref{eq:approx-nac}) differs a little from that given in Ref.~\cite{DRM23}: there we had instead proposed to approximately weigh the doubles contribution with $\sqrt{1 - \alpha_c^2}$, where $\alpha_c$ is a spatially-independent approximation to $\sqrt{a_{{\sss S},3}^{-1} a_c}$ with $a_c(\br,\br') = \bra{\Psi_0}\hat{n}(\br)\ket{\Psi_c}\bra{\Psi_0}\hat{n}(\br')\ket{\Psi_c}$
and likewise $a_{{\sss S},3}$ is the product of the KS transition densities where we use the subscript $3$ to denote the KS state that the TDDFT state $c$ is dominated by, but the weights in Eq.~(\ref{eq:approx-nac}) are better justified in the weak-interaction limit. 
In the specific application to the LiH transition dipole in Ref.~\cite{DRM23}, reproduced in the Supplementary Material here, only the second term was computed (with $G_c = 1$), since the underlying linear response TDDFT calculation was done in the adiabatic approximation over all frequencies,
so would not detect any double-excitation contribution. In reality, a KS double-excitation lies near $\Omega_c$ (see Supplementary Material and also Fig.~\ref{fig:LiH_DSMA} shortly). 
The approximation of Ref.~\cite{DRM23} has an overall trend that follows the exact results,  but becomes an overestimate to the left of the resonance region while underestimating it to the right. 

\begin{figure}[h!]
\includegraphics[width=0.5\textwidth]{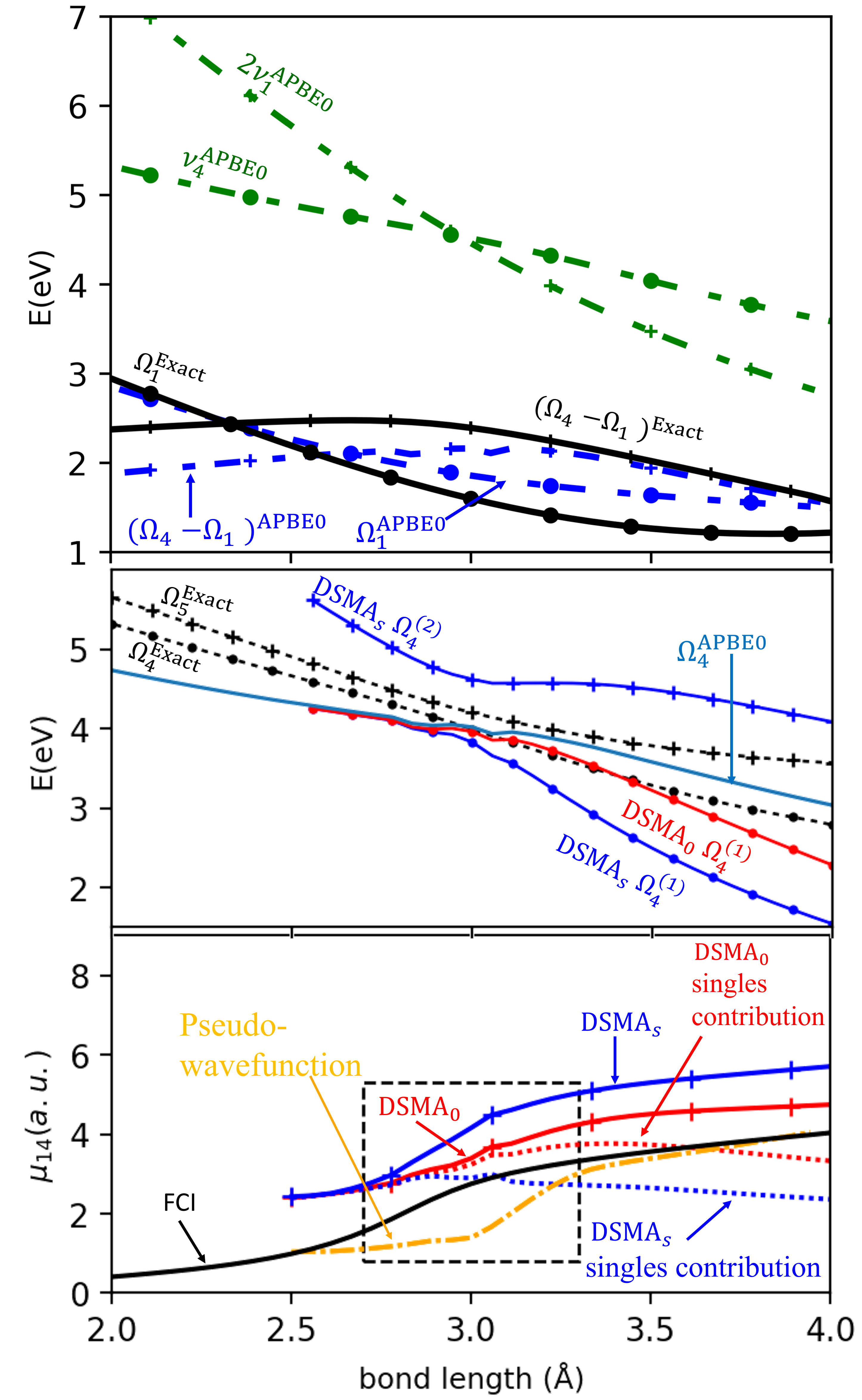} 
\caption{LiH frequencies and $\mu_{14}$. Upper panel: The 4th APBE0 KS excitation frequency, $\nu_4$ is shown intersecting with the KS double excitation, $2\nu_1$ (green). Also shown is the intersection of the 1st singlet excitation frequency, $\Omega_1$ with the difference of the 4th and the 1st excitation frequencies, $\Omega_4-\Omega_1$, computed using APBE0 (blue), and FCI (black), both within aug-cc-pVDZ basis. Middle panel: Excitation frequencies $\Omega_4$ and $\Omega_5$ from FCI (black dashed), $\Omega_4$ (APBE0, light blue) and the two states obtained by applying DSMA$\s$ (blue) and DSMA$_0$ (red) (for which the upper state is out of the scale of the figure).  Lower panel: Transition dipole moment $\mu_{14}$ given by FCI (black), pseudo-wavefunction approximation (orange), DSMA$_0$ (red) and DSMA$_s$ (blue). Also shown is the contribution coming from the 2nd term in Eq.~\ref{eq:approx-nac} for DSMA$_0$ (red dotted) and DSMA$_s$ (blue dotted).}
\label{fig:LiH_DSMA}
\end{figure}

We now use DSMA to compute the proper weight $G_c$ associated with the first and second terms in Eq.~(\ref{eq:approx-nac}) for the transition dipole moment between the 1st and 4th singlet excited states in LiH, $\mu_{14}=\langle \Psi_1\vert\hat{x}\vert \Psi_4\rangle=\int x n_{ca}^{\rm App}(\br) d \br$, where $x$ is along the internuclear axis.  In Fig.~\ref{fig:LiH_DSMA} all calculations are done in the aug-cc-pVDZ basis set using PySCF \cite{pyscf}, while the pseudo-wavefunction calculation is performed with the Turbomole package~\cite{turbomole}~\footnote{{ The pseudo-wavefunction result was obtained using PBE0/aug-cc-pVDZ within the Turbomole package. To get smooth results in this basis, the computational parameters used were: SCF convergence threshold $< 10^{-12} $; TDDFT residual threshold $< 10^{-9}$}}. 
As demonstrated in the Supplementary Material, 
both the TDDFT and reference Full Configuration Interaction (FCI) results show significant basis-set dependence; the transition dipole moments more so than the excitation energies, but convergence appears to be reached with the aug-cc-pVDZ basis. 
The top panel shows the intersection of the energy of the KS double-excitation $2\nu_1^{\rm APBE0}$ with the 4th KS excitation frequency, justifying the need for the DSMA approach. The black lines demonstrate the ``resonance condition" for FCI, $\Omega_4 = 2\Omega_1$, while near the intersection of the corresponding curves for APBE0 (blue lines), $\mu_{14}^{\rm APBE0}$ diverges (see Refs.~\cite{PRF16,DRM23} and Supplementary Material). The middle panel shows the 4th and 5th singlet excitation energies of the exact, and the one APBE0 excitation in this frequency-range, consistent with its blindness to the KS double-excitation in the top panel. Applying our DSMA to this 4th state, we retrieve two states, $\Omega_{4}^{(1)}$ and $\Omega_{4}^{(2)}$, but with some error compared to the exact; both DSMA$\s$ and DSMA$_a$ overshoot the splitting between the two states, with the upper level of the latter being off the scale of the graph. 
The lowest panel compares $\mu_{14}$ computed from Eq.~(\ref{eq:approx-nac}) ($\Omega_c= \Omega_{4}^{(1)})$ using $G_c$ obtained from DSMA$_0$ and DSMA$\s$, against the reference FCI and the pseudo-wavefunction approximation~\cite{LL14,ZH15,OAS15,OBFS15,PRF16}. The latter is often used to tame the divergence of the raw adiabatic transition dipole, but is somewhat {\it ad hoc} in that its underlying kernel structure is not yet known. 
In the indicated region of strong mixing between the single and double excitation $R\sim[2.6\mathrm{\mathring{A}}-3.4\mathrm{\mathring{A}}]$, although they overestimate the dipole, DSMA$\s$ and especially DSMA$_0$ both
have a better trend as a function of $R$ and are closer to the FCI reference than the  pseudo-wavefunction approach. The dotted curves indicate the importance of the double-excitation contribution.

By providing oscillator strengths and excitation energies of states of double-excitation character, the new kernel presented here improves the reliability of TDDFT for linear response spectra, as well as for quadratic response properties where it provides an otherwise missing contribution to the excited-to-excited state transition densities. 
 Future work involves extensive tests on a range of systems including to cases when more than one single-excitation couples strongly with a double-excitation. Also, ramifications of Eq.~(\ref{eq:casida_os}) outside the problem of double excitations will be explored. For example, calculating oscillator strengths using an orbital-dependent functional (exact exchange, hybrids, meta-GGAs) within pure DFT imply a frequency-dependence and  an analogous formula for the generalized KS framework may be needed.

\acknowledgments{
We warmly dedicate this article to John Perdew on his 80th birthday,  as we continue to be guided by his development of approximations that yield "almost the right answer for almost the right reason for almost the right price"~\cite{KLP23}, the success of which in the field of materials science and quantum chemistry speaks for itself. We hope that this contribution is a step in this direction for the problem of oscillator strengths and excited state couplings with states of double-excitation character, and that John will enjoy reading it!
We are very grateful to Shane Parker for providing us with the pseudowavefunction results, and to him and Filip Furche for useful discussions. 
Financial support from the National Science Foundation Award CHE-2154829 is gratefully acknowledged. 
}

\setcounter{figure}{0}
\setcounter{equation}{0}
\onecolumngrid
\pagenumbering{arabic}
\renewcommand{\thepage}{S\arabic{page}}
\renewcommand{\thefigure}{S\arabic{figure}}
\section{Three Flavors of Dressed Single Pole Approximation (DSPA)}
\begin{figure}[h!]
\includegraphics[width=0.5\textwidth]{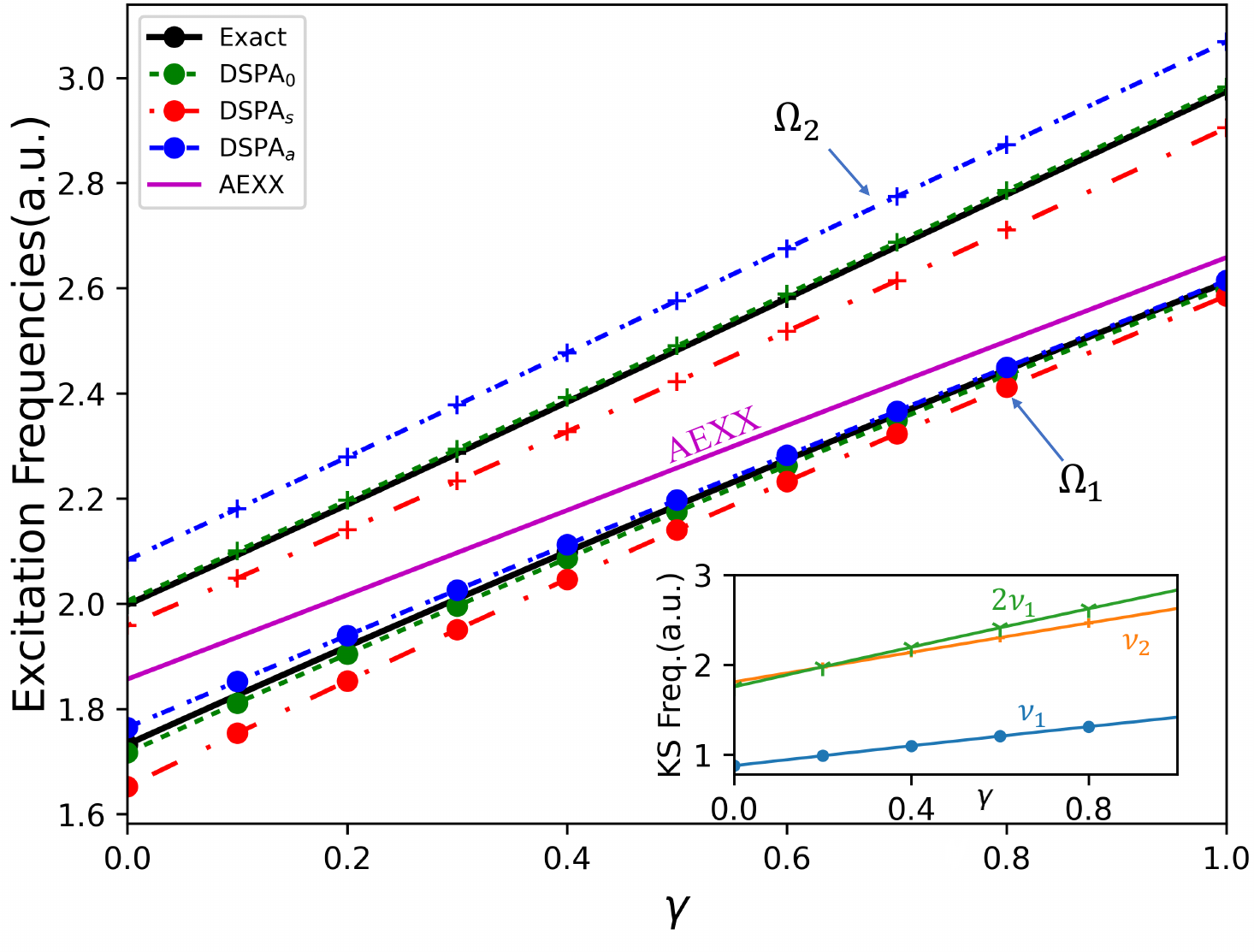} 
 \caption{Two electron model: Exact, AEXX, and the three flavors of DSPA frequencies in the second multiplet, as a function of $\gamma$. The inset shows the KS excitation energy and twice the KS single excitation energy of the first multiplet. The energies are very similar to those given by corresponding DSMA flavors.
  }
\label{fig:DSPA_freq}
\end{figure}

\begin{figure}[h!]
\includegraphics[width=0.5\textwidth]{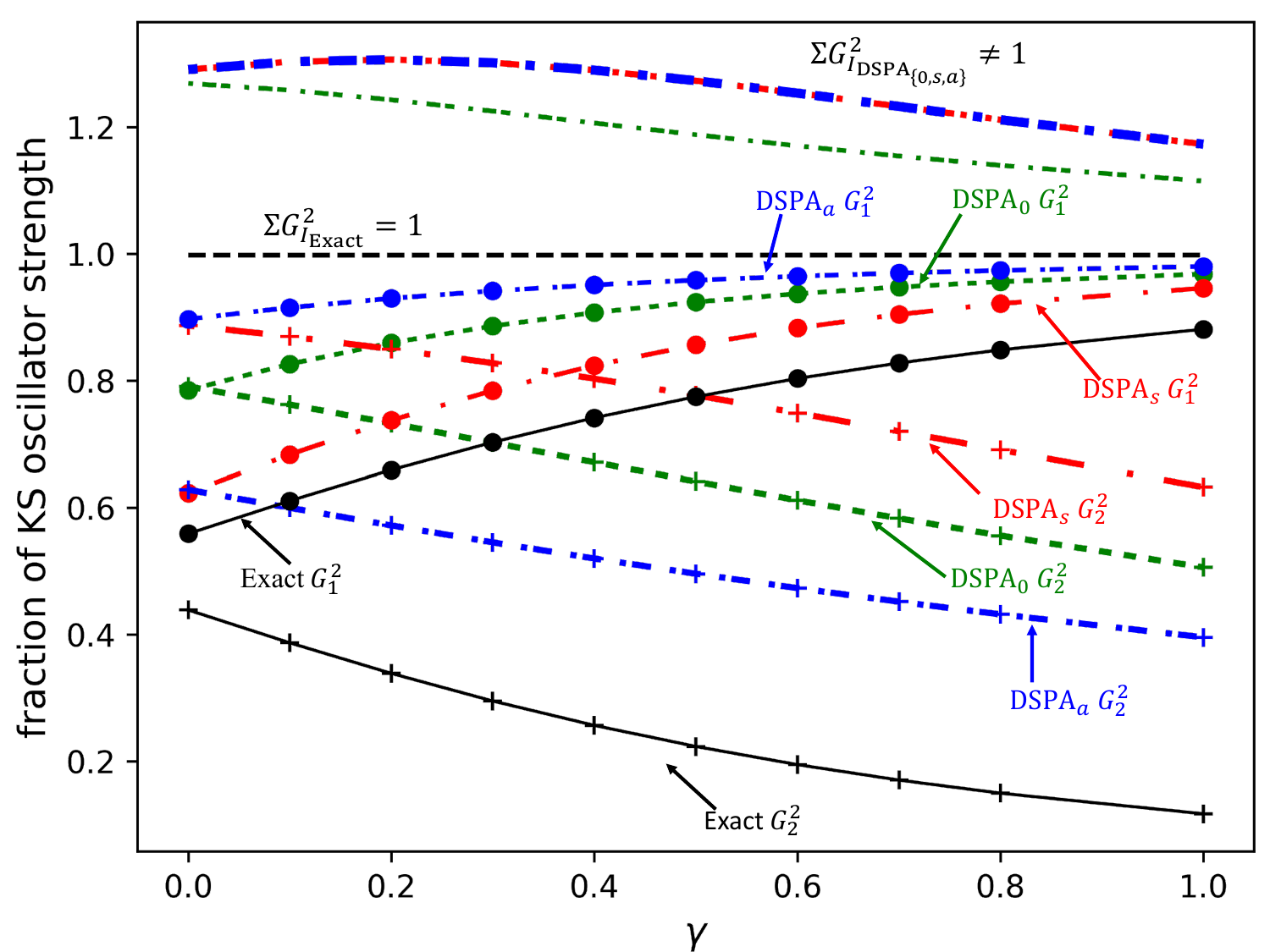} 
 \caption{The fraction of KS oscillator strengths, $G_I^2$ in the exact and the three DSPA calculations. Also shown is the violation of the sum-rule $G_1^2 + G_2^2=1$ by all flavors of DSPA compared to the exact that satisfies the rule.}
\label{fig:DSPA_osc}
\end{figure}

The following are the three flavors of DSPA which were obtained by using the same substitutions as the ones used to get the three DSMA flavors in the main text:
\begin{equation}
f\Hxcqq^{\mathrm{DSPA_0}}(\omega) = f\Hxcqq^{A}  + \frac{H_{qd}^2/2}{\omega - (H_{dd} - H_{00})}
\end{equation}
\begin{equation}
f\Hxcqq^{\mathrm{DSPAs}}(\omega) = f\Hxcqq^{A} + \frac{H_{qd}^{2}/2}{\omega - \nu_d}
\end{equation}
\begin{equation}
f\Hxcqq^{\mathrm{DSPAa}}(\omega) = f\Hxcqq^{A} + \frac{H_{qd}^{2}/2}{\omega - (\Omega_{s1}^A + \Omega_{s2}^A)}
\end{equation}
where symbols have the same meaning as in the paper. Comparing Fig. 1 from the main text to Fig.~\ref{fig:DSPA_freq} and Fig.~\ref{fig:DSPA_osc} we observe that in terms of excited state frequencies, the corresponding DSPA and DSMA flavors are comparable, but DSMA is clearly superior to DSPA for the determination of oscillator strengths.
\section{LiH: Basis Set Dependence of FCI Results and Transition Dipole Moment from Quadratic Response TDDFT}
\begin{figure}
\includegraphics[width=0.5\textwidth]{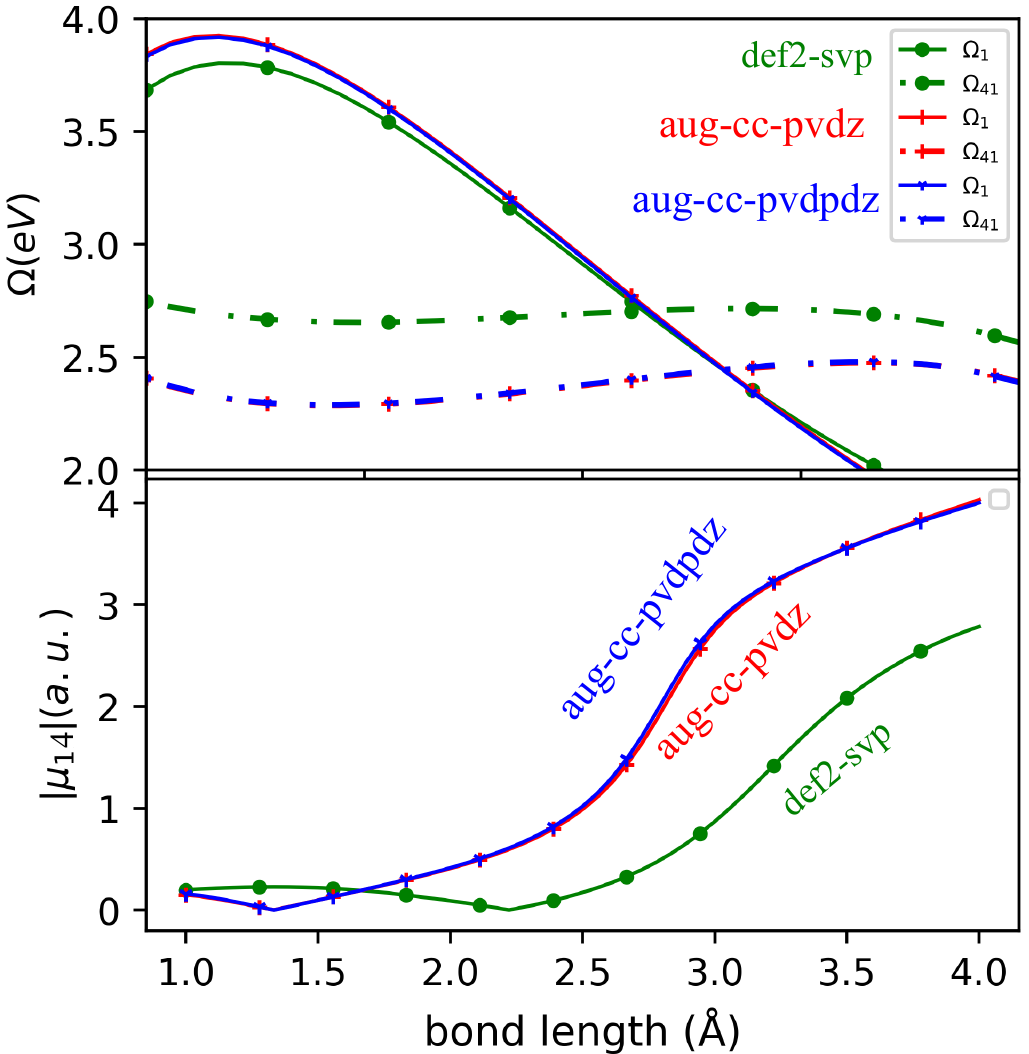}
    \caption{Top Panel: FCI calculations of the frequency of the 1st excited state  $\Omega_1$ and the difference between the 4th and the 1st excited state frequencies $\Omega_4-\Omega_1$ in the labeled basis sets.
    Lower Panel: the corresponding transition dipole moments between the 1st and the 4th excited states.}

    \label{fig:basis_set}
\end{figure}

\begin{figure}[h]
    \includegraphics[width=0.5\textwidth]{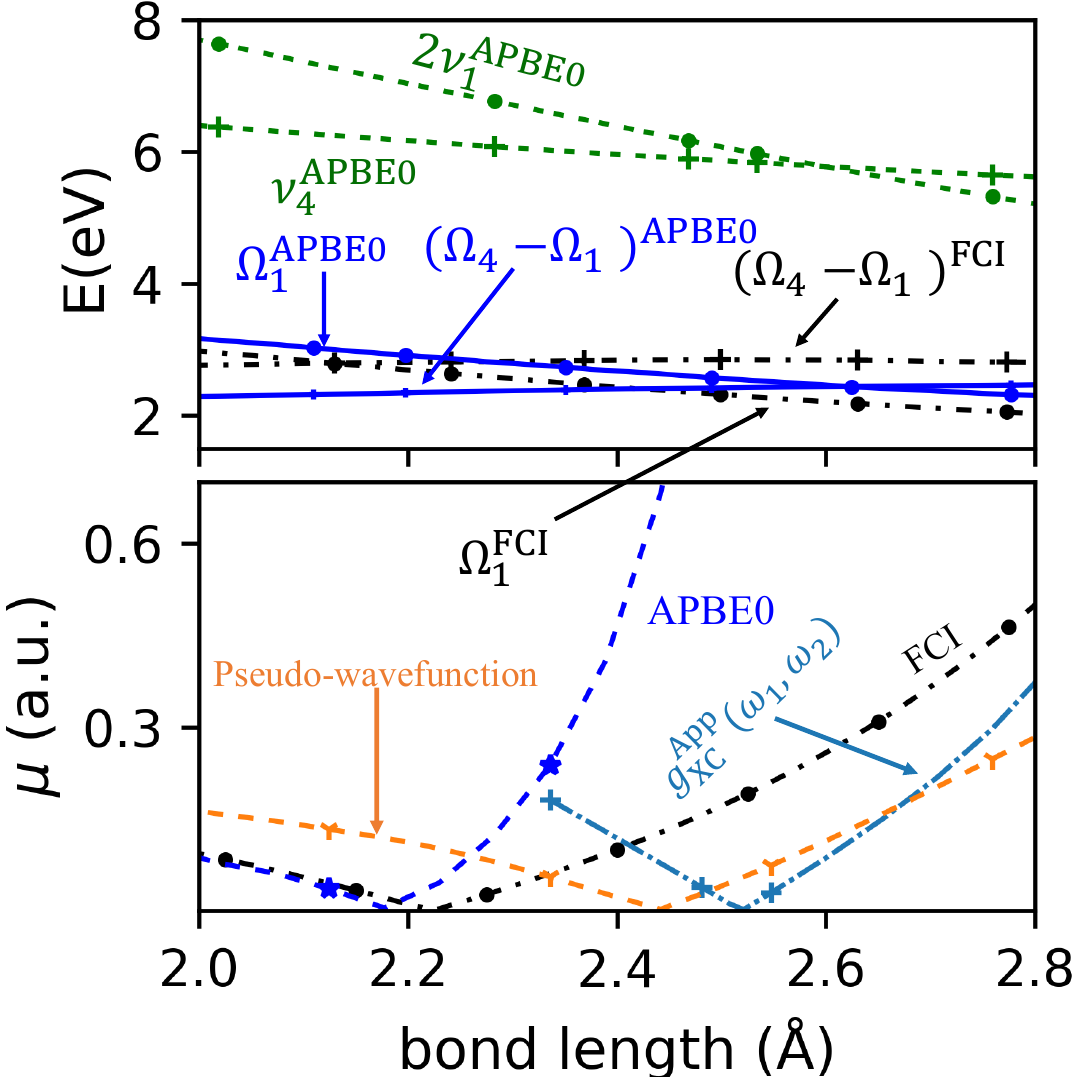}
    \caption{Transition dipole moment $\mu_{14}$ in the LiH molecule as a function of internuclear separation $R$. Upper panel shows TDDFT excitation frequency of the 1st excited state, $\Omega_1$, and the difference in frequencies of the 4th and the 1st excited state, $\Omega_4 - \Omega_1$; The Kohn-Sham, frequency $2\nu_1$ intersecting with $\nu_4$ calculated with APBE0 within def2-SVP basis~\cite{DRM23,PRF16}. In black dashed are the exact frequencies of Full CI in this basis. Lower panel shows $\mu_{14}$ calculated in the various methods indicated in the same basis, compared to the exact  $\mu_{14}$}
    \label{fig:LiH}
\end{figure}

The full configuration interaction (FCI) results for the frequencies and transition dipole moment $\mu_{14}$ between the 1st and the 4th singlet excited state of LiH molecule in the main text appear quite different from those shown in Refs.~\cite{DRM23,PRF16}, due to using a different basis set. The TDDFT results for the dipole moment, using adiabatic TDDFT, pseudowavefunction, or the frequency-dependent g$\xc$ kernel from Ref.~\cite{DRM23} (Eq. (15) of main text) are all also quite sensitive to basis set, more so than the TDDFT energies. This should be borne in mind when comparing against a reference: given the different basis set sensitivities of different methods, thought needs to be given to which basis a FCI reference calculation should be compared to when gauging the accuracy of a  TDDFT approximation performed in a given basis. To illustrate this, we compare the excited state frequencies and transition dipole moments given by the FCI within different basis sets in Fig.~\ref{fig:basis_set}. Going from the def2-svp basis to the augmented Dunning basis sets aug-cc-pVDZ and aug-cc-pVDPDZ, the transition dipole moment shows more sensitivity than the excited state frequencies. Using progressively bigger basis sets in the augmented Dunning basis set series results in convergence for $\mu_{14}$ which we use as reference in the main paper.
Next, we reproduce the results of Ref.~\cite{DRM23}  which used the def2-SVP basis,  to compare with the figure in the main text of the present paper that uses the aug-cc-pVDZ basis. Figure~\ref{fig:LiH} shows the transition dipole moment $\mu_{14}$ between the 1st and 4th excited states of LiH molecule in the A$_1$ irreducible representation of $C_{2v}$ point group symmetry. The dipole  $\mu_{14}=\langle \Psi_1\vert\hat{x} \vert \Psi_4\rangle=\int x  n_{ca}(\br) d \br$  is computed along the internuclear axis $\vec{x}$, as a function of bond-length using the adiabatic approximation,  APBE0 within quadratic response TDDFT, the pseudo-wavefunction approach and using Eq.~(15) of the main text by taking $G_c = 1$, as was done in Ref.~\cite{DRM23}. All calculations are done with def2-SVP basis set, using PySCF \cite{pyscf} with the exception of pseudo-wavefunction which is done using the Turbomole package~\cite{turbomole}. As can be seen, in the region around 2.6$\mathrm{\mathring{A}}$ the 4th excited state frequency is close to twice the 1st excited state frequency (while in the larger aug-cc-pVDZ basis shown in the main text, the intersection appears closer to  2.6$\mathrm{\mathring{A}}$). Compared to the FCI $\mu_{14}$ shown in the lower panel, ABPE0 suffers from the expected divergence while the approximation, which lacks the proper oscillator strength fractions, effectively mitigates the divergence observed in the transition dipole moment obtained from the adiabatic approximation in the region where the resonance condition $\Omega_b = \Omega_c - \Omega_a$ is satisfied. We see that the approximation aligns with the overall pattern observed in the exact results, however, it tends to overestimate the values to the far left of the resonance region while underestimating it to the right. 
\bibliography{DDSMA.bib}


\end{document}